\documentclass[letterpaper, 10 pt, conference]{ieeeconf}  

\IEEEoverridecommandlockouts                              
\usepackage{graphicx} 
\usepackage{amsmath} 
\usepackage{amssymb}  
\usepackage{multirow}
\usepackage{xcolor}
\usepackage{todonotes}
\usepackage{url}
\usepackage{array}
\newcolumntype{C}[1]{>{\raggedright\let\newline\\\arraybackslash\hspace{0pt}}m{#1}}

\title{\LARGE \bf
Optimising Lockdown Policies for Epidemic Control using Reinforcement Learning \\ {\normalsize{Summary, mathematical formulation, and simulation results for India}}
}

\author{Harshad Khadilkar$^*$, Tanuja Ganu$^+$, Deva P Seetharam$^\#$\\$^*$TCS Research, $^+$Microsoft Research, $^\#$Independent consultant
\thanks{{harshad.khadilkar@tcs.com}, {harshadk@iitb.ac.in}, {tanuja.ganu@microsoft.com}, {deva.seetharam@gmail.com}}%
}

\begin{document}

\maketitle
\thispagestyle{empty}
\pagestyle{empty}

\begin{abstract}

In the context of the ongoing Covid-19 pandemic, several reports and studies have attempted to model and predict the spread of the disease. There is also intense debate about policies for limiting the damage, both to health and to the economy. On the one hand, the health and safety of the population is the principal consideration for most countries. On the other hand, we cannot ignore the potential for long-term economic damage caused by strict nation-wide lockdowns. In this working paper, we present a quantitative way to compute lockdown decisions for individual cities or regions, while balancing health and economic considerations. Furthermore, these policies are \textit{learnt} automatically by the proposed algorithm, as a function of disease parameters (infectiousness, gestation period, duration of symptoms, probability of death) and population characteristics (density, movement propensity). We account for realistic considerations such as imperfect lockdowns, and show that the policy obtained using reinforcement learning is a viable quantitative approach towards lockdowns.

\end{abstract}


\section{Executive Summary} \label{sec:summary}

\textbf{Important note:} \textit{While the initial version of the model used randomly generated network data, this version is based on population and case data from public sources in India. \textbf{However, the results are generated using approximate modelling techniques, and should not be considered quantitatively exact, and nor should they be used without context for designing public policy.} We wish to disseminate this work as soon as possible, in the context of the current Covid-19 urgency. We will continue to add more detailed description and literature review in stages.}\\

The goal of this section is to present a non-technical summary of the methodology and the results of this study. More detailed mathematical description is available in Section \ref{sec:math}.

\subsection{Contributions}

The key contributions of this study are as follows.
\begin{enumerate}
\item We provide a methodology to compute the optimal lockdown/release policy for each node in a network, given disease characteristics and network properties.
\item This optimal policy is automatically learnt by the algorithm, using a technique known as reinforcement learning \cite{Sutton:2012}. No explicit optimisation knowledge is required on the part of the user.
\item The policy is explicitly cognizant of health and economic costs, with the weight on each factor specified by the user's preferences.
\item The same algorithm can be used to compute the policy for any changes in (i) network data, (ii) disease parameters, and (iii) cost definitions. Only the relevant input values need to be updated by the user.
\item The computed policies are explainable, which means that the users can ask questions about the interventions and the reasons behind each decision. In that sense, the optimisation algorithm is \textit{not a black box}.
\end{enumerate}

\subsection{Disclaimers}

Before proceeding further, we wish to make explicit the limitations of this study.
\begin{enumerate}
\item None of the authors are experts on communicable diseases. We have used disease models from available literature \cite{perez2009agent}, which has a fair degree of consensus. The accuracy of that framework in the context of Covid-19 is as yet unknown. However, we do model some of its specialities such as asymptomatic carriers, and the fact that transmission can happen from people who are exposed to the virus but not yet showing symptoms.
\item The network model used in this study is basic and not very sophisticated. It does account for population size and geography, but the effects are at a macro level. We do not model at the level of specific individuals and their movements. 
\item We use publicly available parameters relating to Covid-19 and the district-level network for India, but these are highly approximate. We have tried to match the characteristics of Covid-19 as closely as possible, within the constraints of time and tractability. Further enhancements are possible through tuning of parameters by epidemiological experts.
\item We use sample data from India for illustration, but the results we present need not be quantitatively accurate. As we show in Section \ref{subsec:validation}, the real data shows characteristics that we cannot fully explain with high-level models of the network and epidemiology. However, we believe that the reinforcement learning approach will still work effectively with more detailed network and disease models. This is work in progress.
\end{enumerate}

\subsection{Methodology in brief}

We approach this study as a combination of three steps. These are described below, with details in Section \ref{sec:math}.

\begin{figure}[t]
\centering
\includegraphics[width=0.3\textwidth]{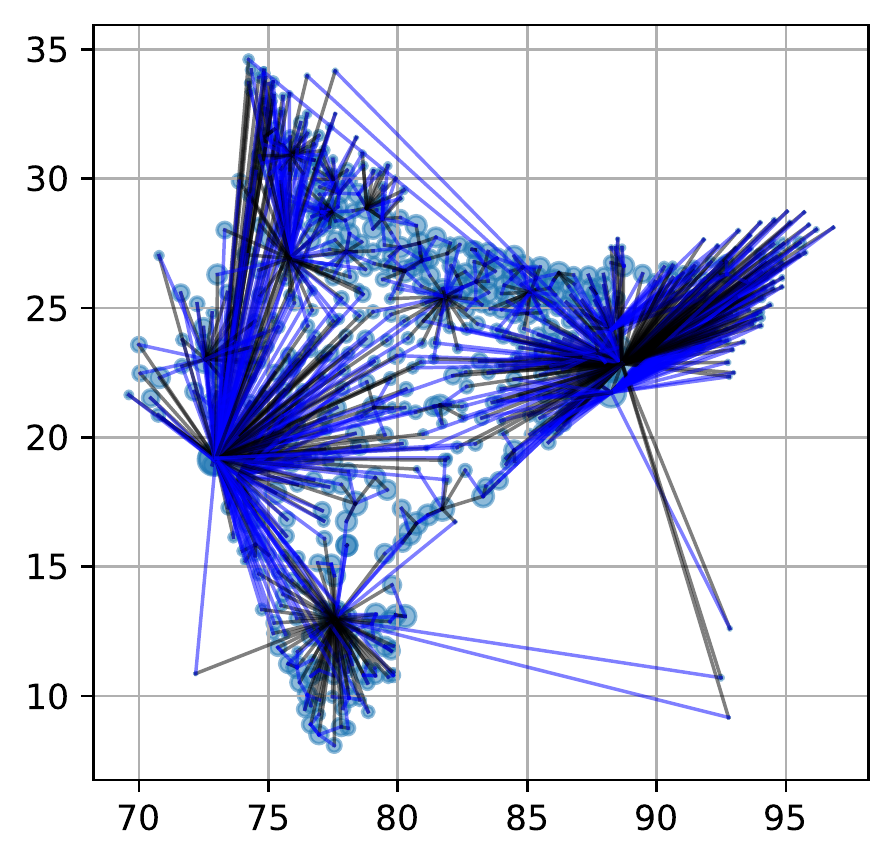}
\caption{Network layout, with size of bubbles indicating population size. Black lines indicate strongest connections for each node, while blue lines indicate the next strongest set of connections. Note that each pair of nodes is potentially linked with each other.}
\label{fig:network}
\end{figure}

\textbf{Obtaining network data for India:}
We use district-level data from India\footnote{https://www.census2011.co.in/district.php, retrieved 04/04/2020}, containing 640 districts (nodes), their positions (latitude and longitude) and the approximate population of each district. Strength of connections between each pair\footnote{Resulting in a fully connected network.} of nodes is (i) directly proportional to the product of the populations of the two nodes, and (ii) inversely proportional to the square root of the distance between them. Prior work on social networks typically uses gravity models \cite{allamanis2012evolution} in which the strength of interaction decreases with the square of the distance. However, these models are for social networks, while the present context is about geographical networks (which can be strongly connected across large distances by air travel). The resulting network structure is shown in Fig. \ref{fig:network}. Note that the figure only shows the two strongest connections for each node, but that all pairs are theoretically connected.

\begin{figure}[t]
\centering
\includegraphics[width=0.48\textwidth]{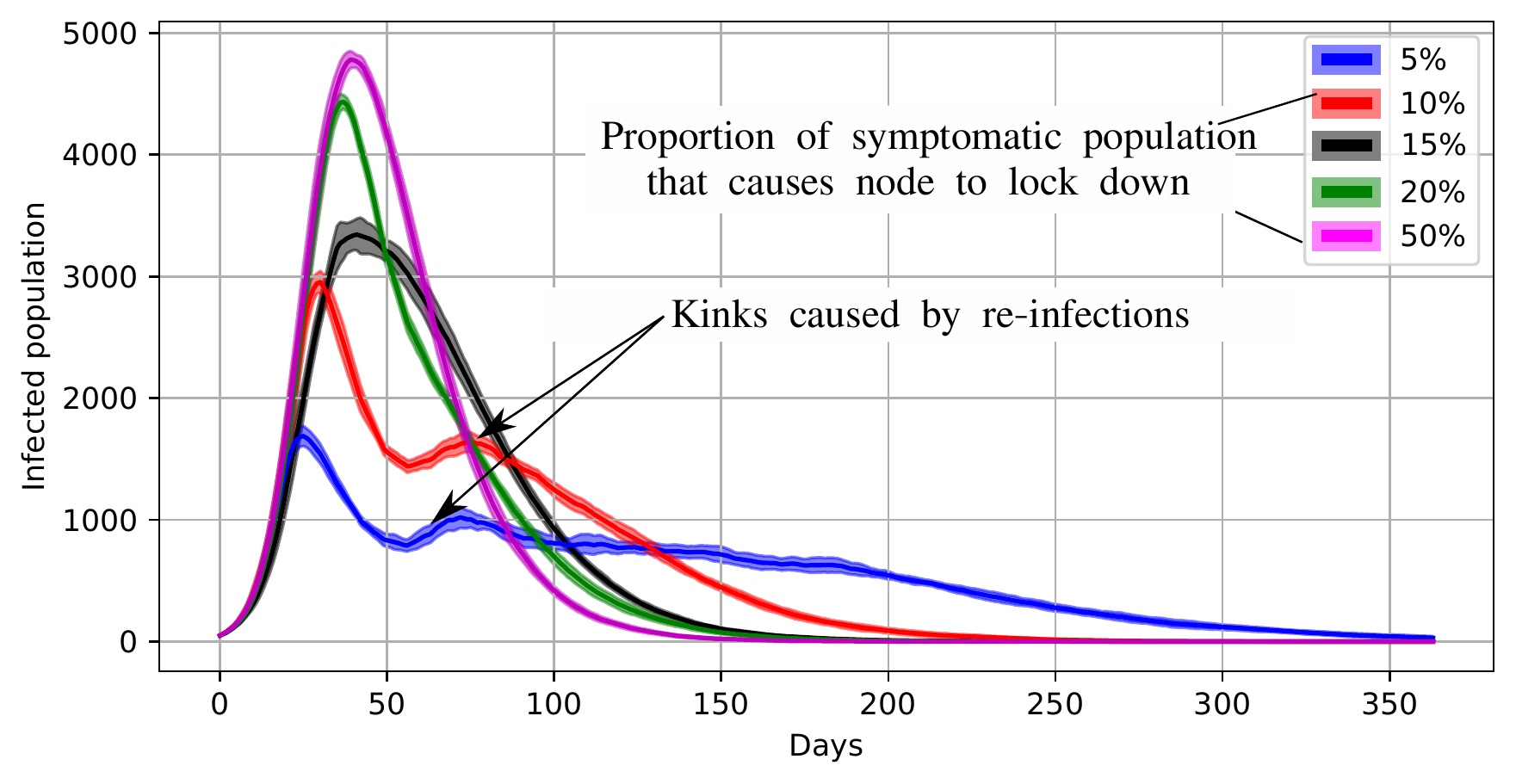}
\caption{Evolution of infection rates for different baseline policies, in the case of a randomly generated network with a small population.}
\label{fig:sum-randomdata}
\end{figure}
\begin{figure}[t]
\centering
\includegraphics[width=0.48\textwidth]{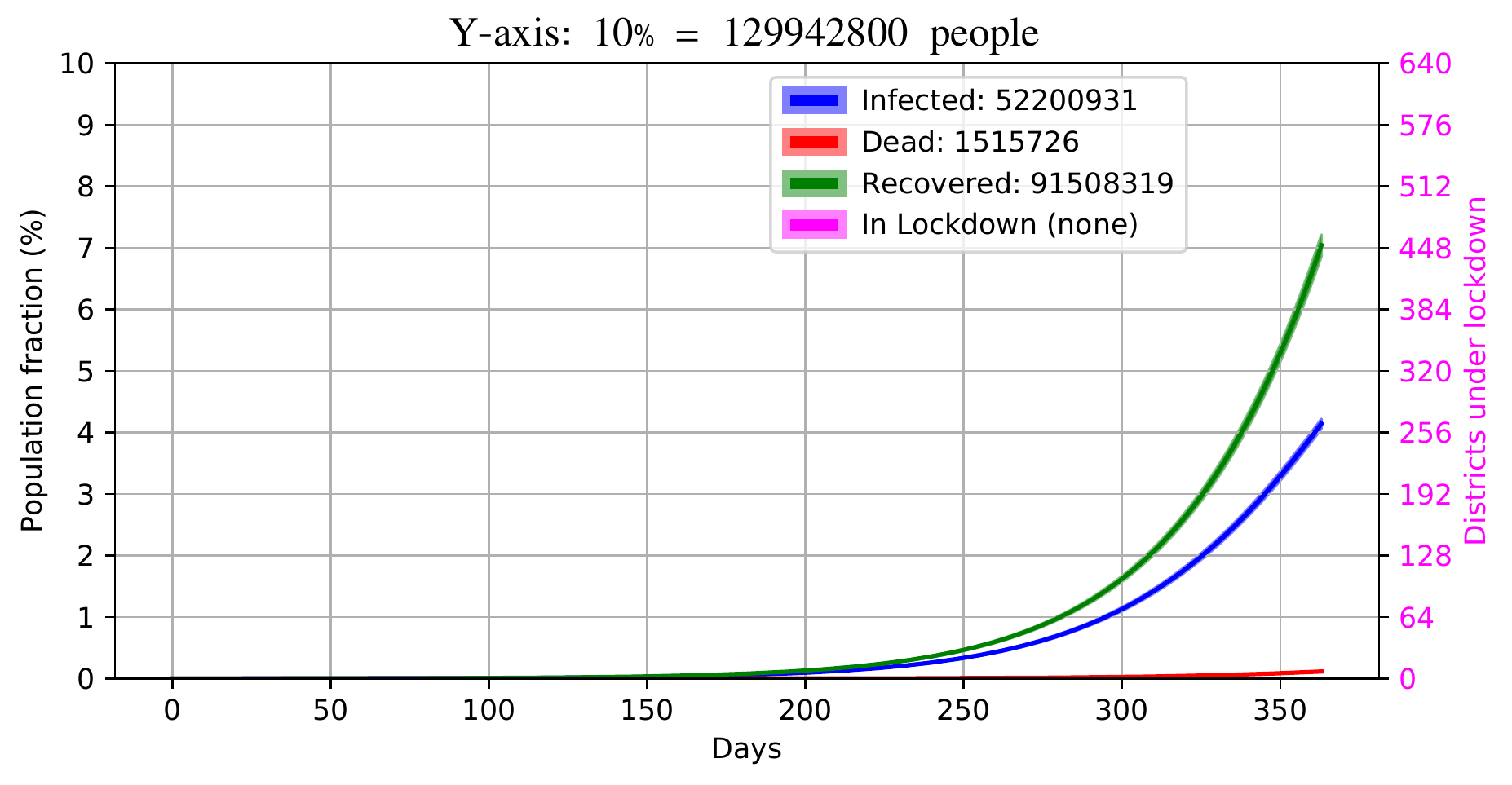}
\caption{Evolution of infection rates if no districts in India were ever locked down, starting from an initial randomly distributed infected population of 5000. Numbers at the end of the year are written in the legend.}
\label{fig:sum-no-100}
\end{figure}

We also model the disease parameters as available online \cite{jung2020real}, for Covid-19. There is a spread in reported numbers across studies, so we use representative values from a cross-section of studies. Specifically, the average incubation period is 10 days, the average infected period is 14 days, exposed people can begin transmitting the disease immediately, only 80\% of people actually show visible symptoms, and the death rate for infected and symptomatic people is 2\%. We also assume that if an exposed or infected person comes into contact with a susceptible person, the transmission probability is 10\%. All the parameters described here are probabilistic, and values for each specific step of the simulation are generated at runtime. We run multiple simulations to get reliable statistics.

\textbf{Effect of lockdown:}
We assume that an open node allows people to travel to/from other open nodes in the network. However, people showing symptoms are not allowed to travel to other nodes (asymptomatic and exposed people can do so). Furthermore, symptomatic people are quarantined within the node, but a small fraction may still circulate within the node by breaking the quarantine. Initial results assume that when a node is locked down, all travel to/from the node is blocked. However, a small fraction of people may still circulate within the node (especially ones who are not showing symptoms). Finer control can be obtained by modulating the severity of lockdown, which we discuss in Section \ref{subsec:finecontrol}.

\textbf{Baseline lockdown policies:} 
The phrase \textit{flattening the curve}\footnote{https://coronavirus.jhu.edu/data/new-cases} has gained popularity in recent weeks. It refers to the fact that social distancing and lockdowns leads to a decrease in the rate of new infections, compared to the sharp peak observed without these measures. Fig. \ref{fig:sum-randomdata} shows this effect for a hypothetical network with 100 nodes and a population of 10000. Each node in the network imposes lockdown when a fixed fraction of the population shows symptoms. The lower the value of this threshold, the flatter the curve (and the longer the epidemic lasts).

We now apply these policies to the India census data. We start all simulations from an initial infected population of 5000 people (out of a total population 1.3 billion), randomly distributed amongst the 640 districts in the India data set. Each node has the option of locking down on any given day; however, once a node is locked down, it must remain so for the next week. The simulation time step is 1 day, and uses the well-known SEIRD model dynamics \cite{perez2009agent} for computing changes of state. The first baseline that we simulate is the \textit{never lockdown} policy, which carries on business as usual. The results are shown in Fig. \ref{fig:sum-no-100}, over a period of 1 year. We only see the exponentially rising portion of the default curve, because of the large population and geographical distribution. Nevertheless, the number of deaths even in this initial rise is clearly unacceptable.

\begin{figure}
\centering
\includegraphics[width=0.48\textwidth]{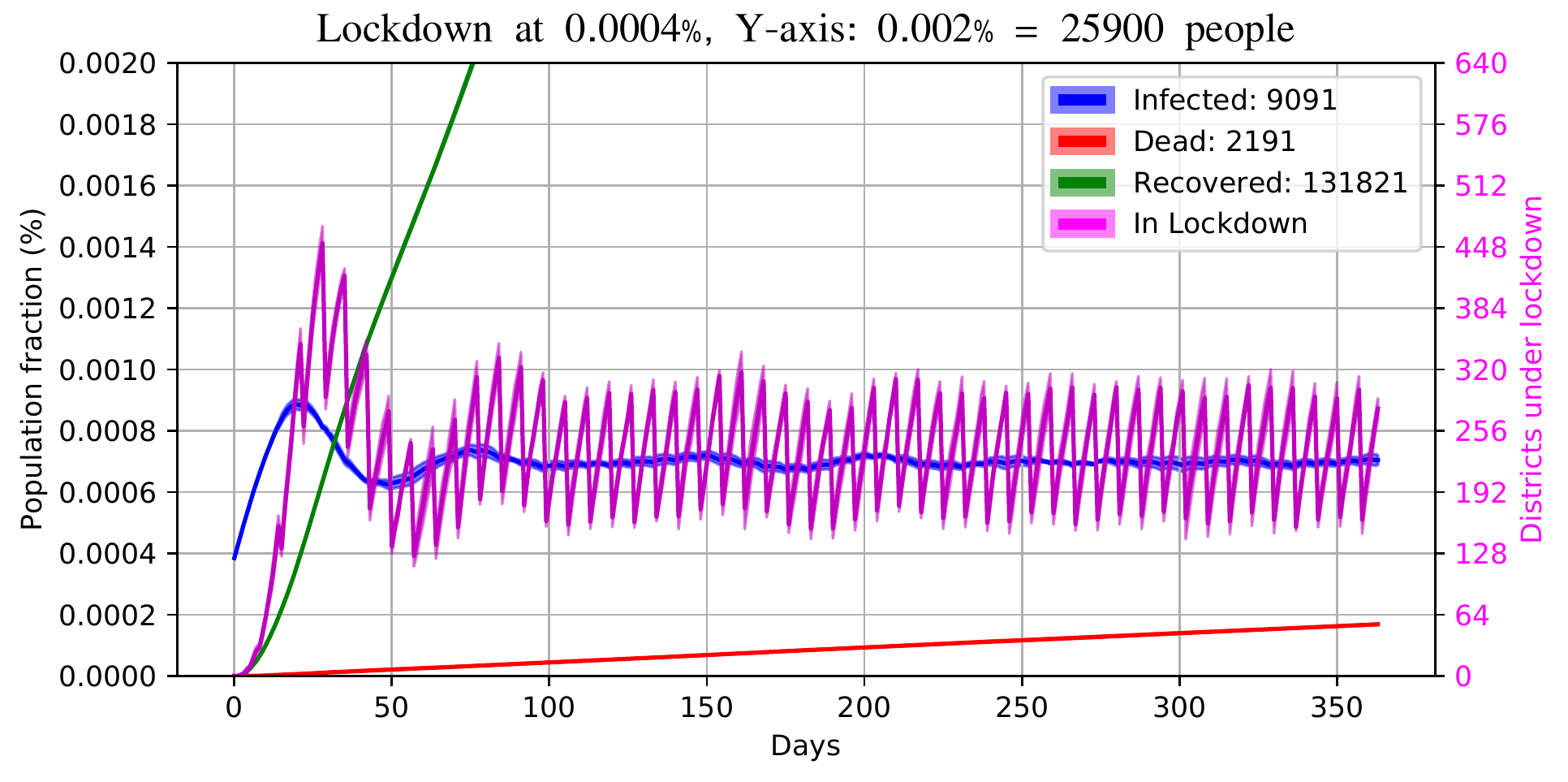}
\vskip-5pt
\caption{Evolution of infection rates when each district is locked down if the symptomatic population exceeds 0.0004\% (4 ppm), and kept open otherwise. The curve is flat, but the infection rate does not decay to 0.}
\label{fig:sum-no-0004}
\end{figure}

The alternative is to impose lockdowns at some threshold, just like the hypothetical network in Fig. \ref{fig:sum-randomdata}. However, given the large population, the threshold for this network must be much smaller. In Fig. \ref{fig:sum-no-0004}, we show the simulation results with districts being locked down at a threshold of 4 symptomatic people per million\footnote{The average population in a district is 2 million.}. The rise of the infection rate is arrested very early in this simulation; however, the number of districts in lockdown fluctuates between 180 and 315 (out of a total of 640) throughout the year. Additionally, the number of infections remains constant at approximately 9000 in the country. Unless pharmaceutical interventions are imposed, it will take a long time for the epidemic to end.
%
%
\begin{figure}
\centering
\includegraphics[width=0.48\textwidth]{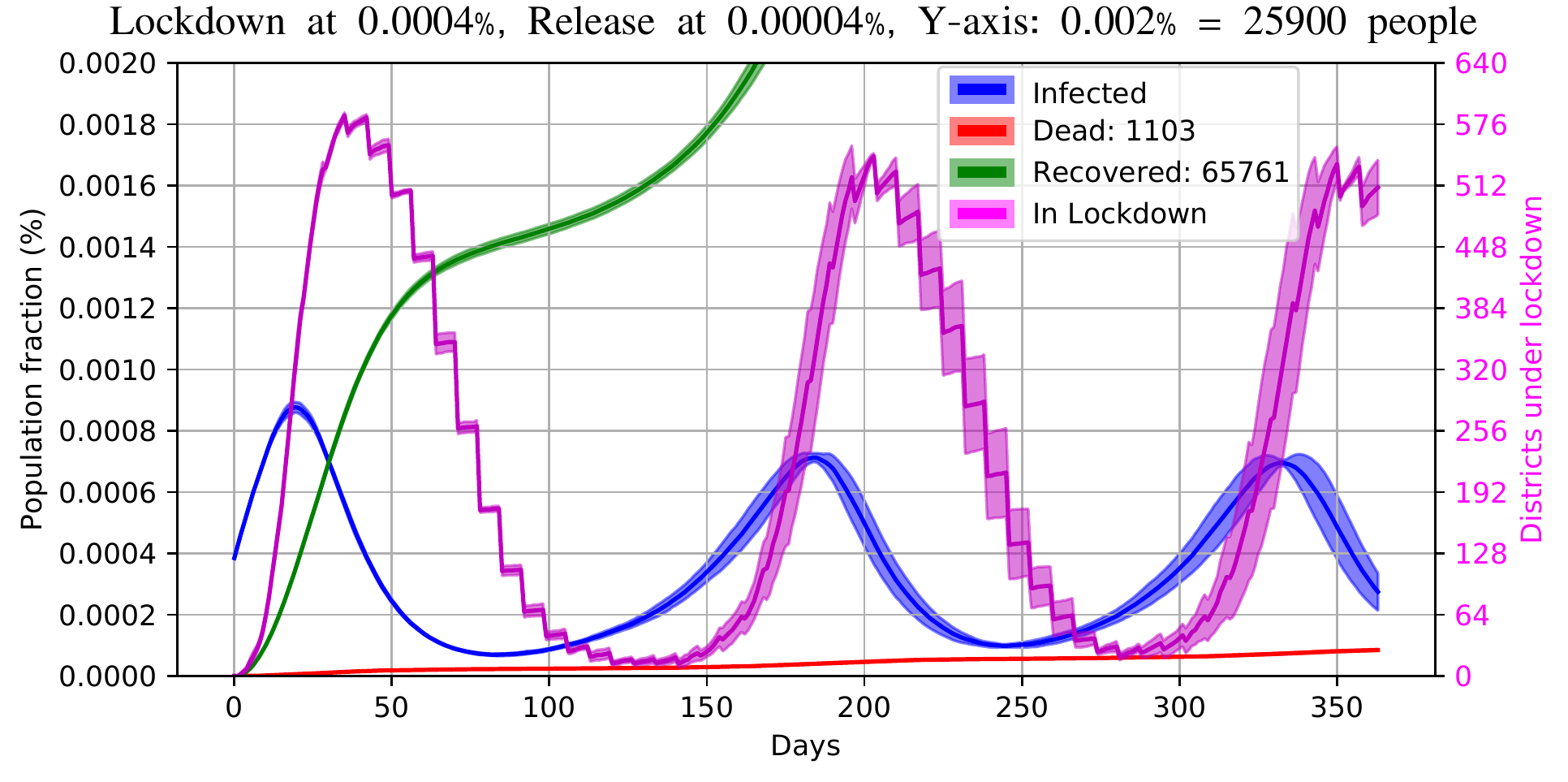}
\vskip-5pt
\caption{Evolution of infection rates when each district is locked down if the symptomatic population exceeds 4 ppm, and released once it falls below 0.4 ppm. Infection re-emerges whenever lockdowns are relaxed.}
\label{fig:sum-hyst}
\end{figure} 

A smoother change in the number of lockdowns is observed in Fig. \ref{fig:sum-hyst}, where the districts impose restrictions at the same threshold (4 symptomatic people per million), but only relax the restrictions when the known infection rate falls below 0.4 people per million. However, we see that the infection re-emerges whenever the restrictions are relaxed. Both Fig. \ref{fig:sum-no-0004} and Fig. \ref{fig:sum-hyst} show a relatively slow fall in infections and fast re-emergence for two reasons: (i) Covid-19 can be transmitted by people who are exposed to it, even before they begin to show symptoms (resulting in a large number of infections even before the first person shows symptoms), and (ii) we assume that lockdowns are leaky, in the sense that a small fraction of people (including an even smaller fraction of technically quarantined people) circulate within nodes even during lockdowns.

Instead of predefined policies, we now describe an approach that \textit{learns} the optimal policy, given the economic and health cost parameters.

\begin{figure}[t]
\centering
\includegraphics[width=0.48\textwidth]{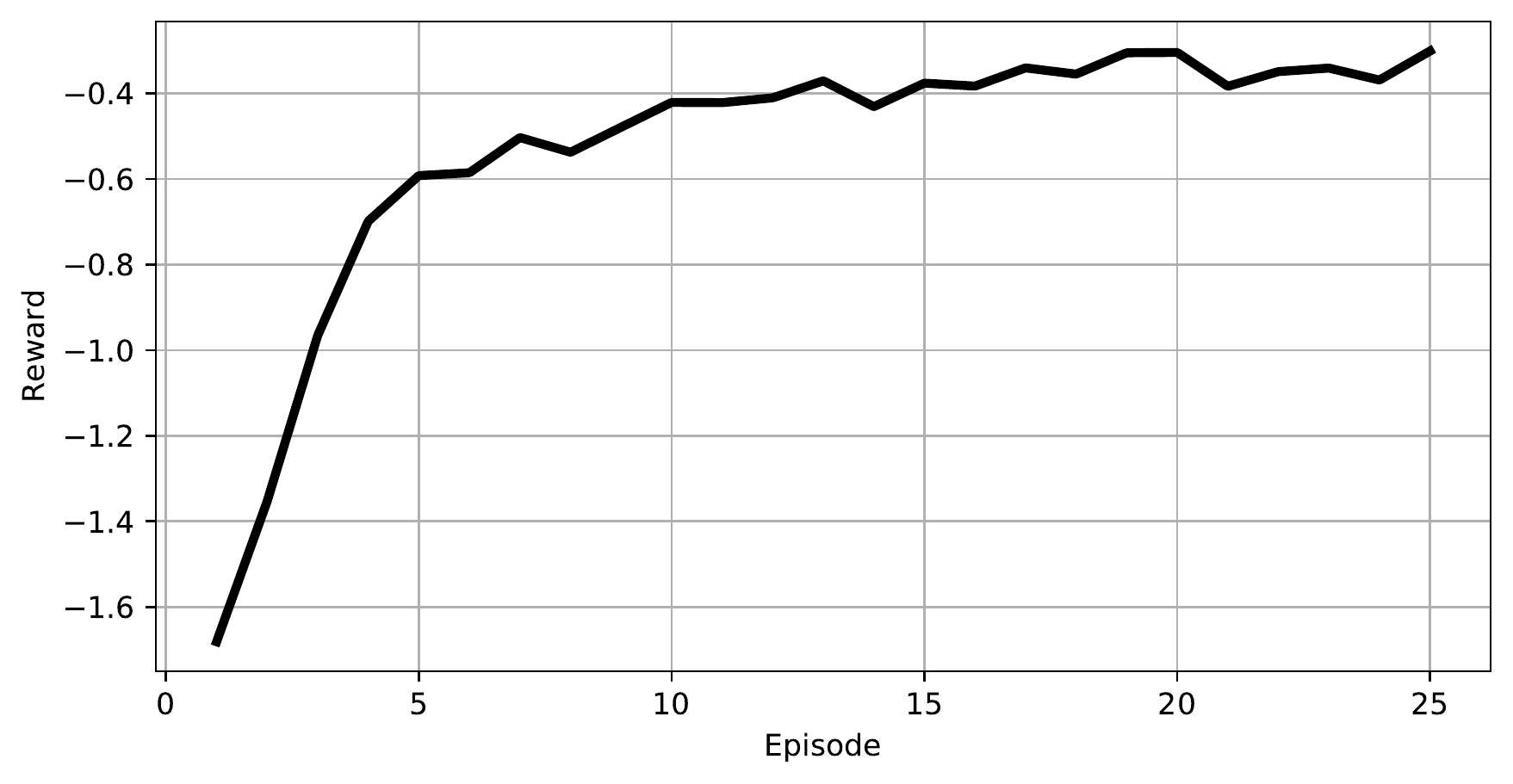}
\vskip-15pt
\caption{Evolution of reward during RL training. Described in detail in Section \ref{sec:math}, the reward is based on number of days of lockdown, number of infections, and number of deaths.}
\label{fig:rew-rl}
\end{figure}

\textbf{Train the reinforcement learning (RL) algorithm:}
RL works by running a large number of simulations of the spread of the disease, while experimenting with different rules for lockdowns. The chief requirement is to quantify the cost of each outcome of the simulation. In this study, each week of lockdown for an average district has the same economic cost as 0.2 infections or 0.1 deaths per million population\footnote{However, these costs are fully tunable}. A \textit{reward} is defined as the negative of these costs (higher the reward, lower the cost). As shown in Fig. \ref{fig:rew-rl}, the RL algorithm is able to improve the reward to a stable level over approximately 25 simulations. 

\begin{figure}[t]
\centering
\includegraphics[width=0.48\textwidth]{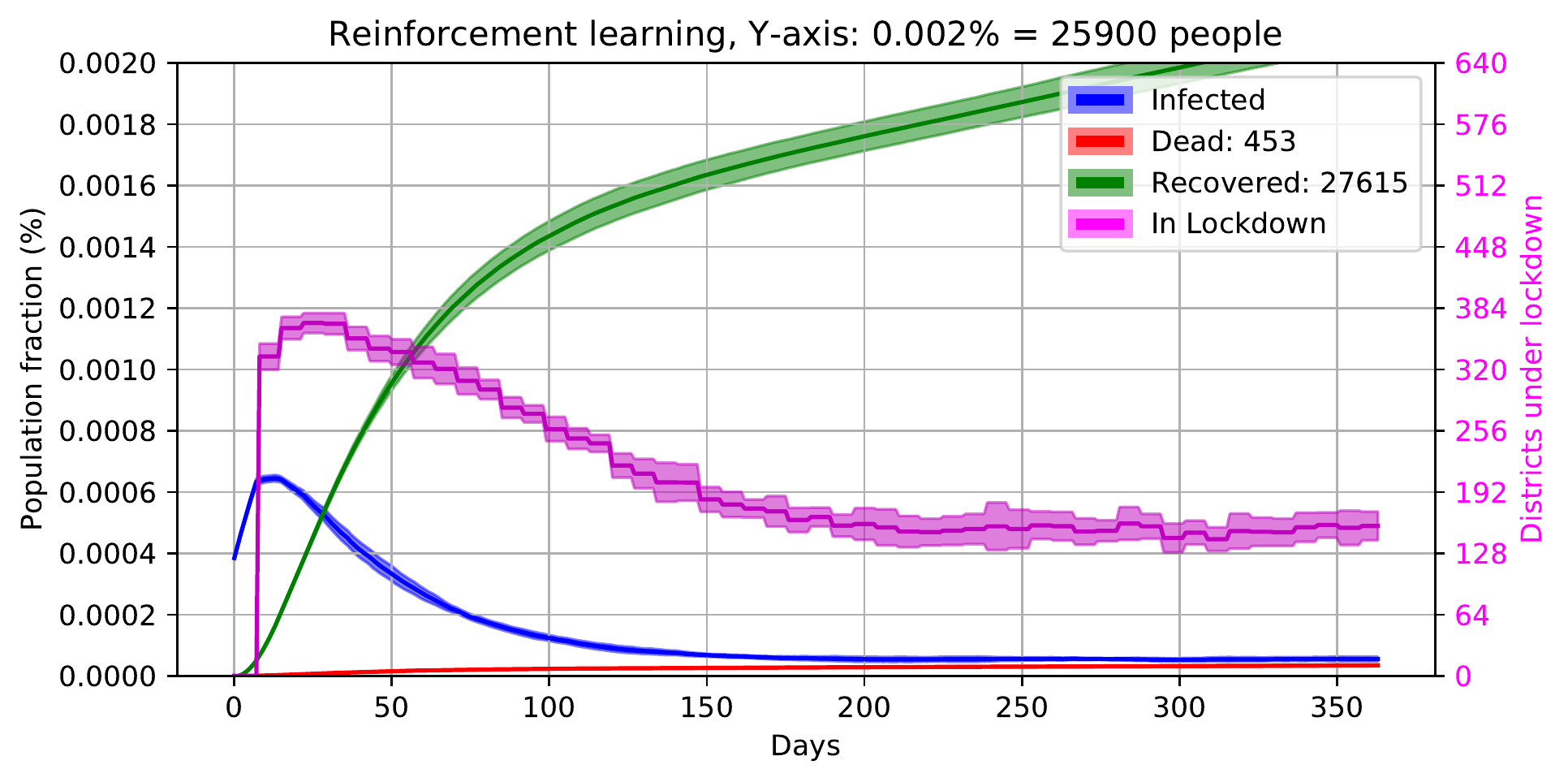}
\vskip-15pt
\caption{Evolution of infection rates when each node is locked down according to the learnt reinforcement learning policy. Instead of a slow imposition of lockdowns, the policy `goes hard and goes early' with a large number of districts, and relaxes lockdowns much slower than the baselines.}
\label{fig:sum-rl}
\end{figure}

\begin{figure}[t]
\centering
\includegraphics[width=0.48\textwidth]{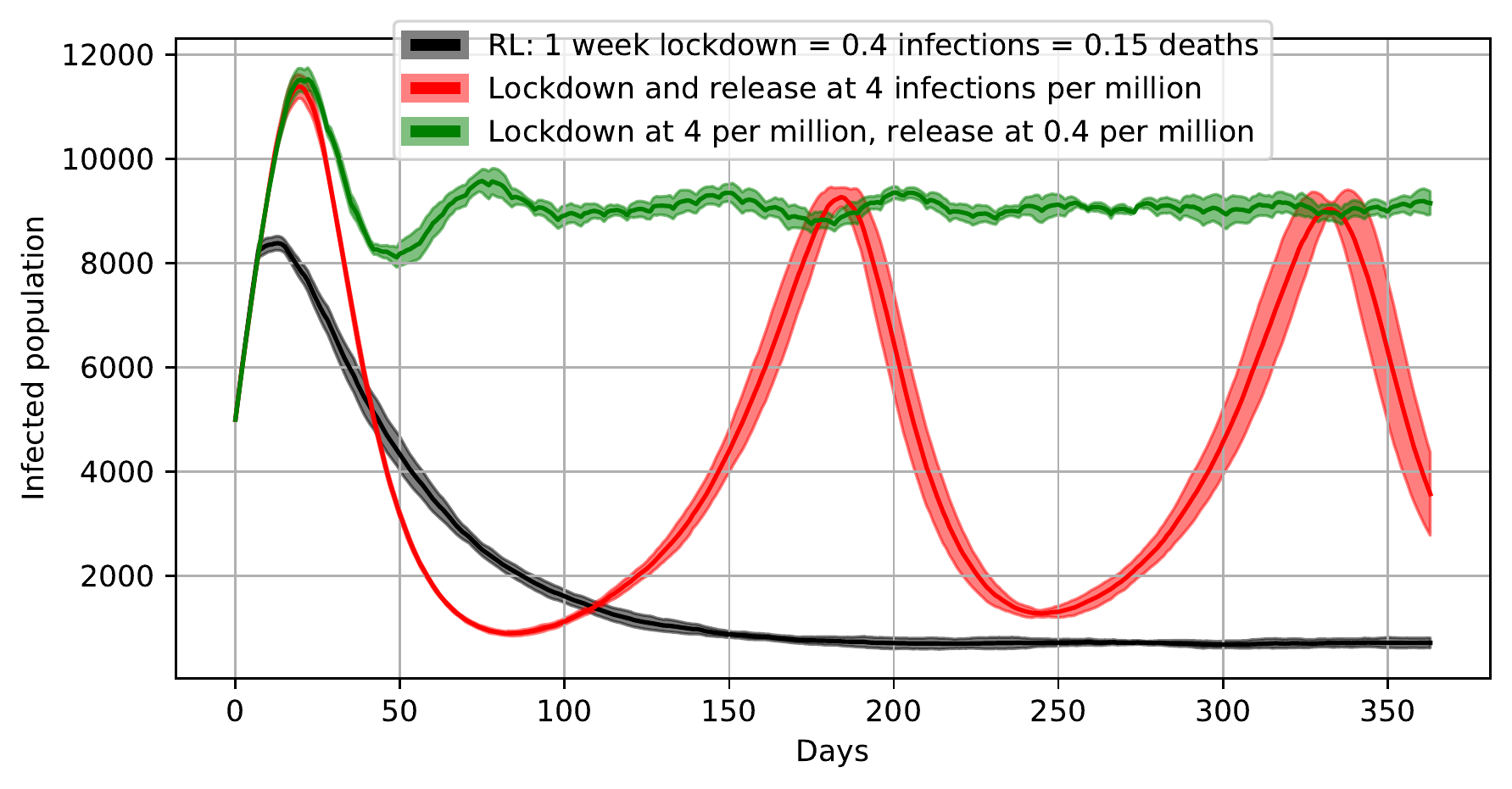}
\vskip-15pt
\caption{Comparison of infection rates for two baselines and the RL policy.}
\label{fig:sum-all}
\end{figure}

The evolution of infection rates in Fig. \ref{fig:sum-rl} shows that the RL policy imposes fewer lockdowns than either of the baseline policies we described before, and yet results in a reduction in infections and fewer deaths. The key difference is that initial lockdowns are imposed much earlier, and simultaneously for a large fraction of at-risk districts. Following this, the relaxation happens much more slowly than for the baselines. The net effect of these decisions is to control the disease early and to keep it localized in the initial districts. The RL policy also uses much more \textit{context} when computing lockdown decisions. While this is described in detail in Section \ref{sec:math}, the intuition is as follows: Not only is it important to look at the number of detected infections within the district, one must look at additional factors such as overall infections in the country, the rates of increase/decrease in recent days, and the size of susceptible population in the district. This makes the policy sensitive to `red flags' such as small breakouts of infection within the node, which causes lockdowns to be enforced sporadically throughout the year.
Finally, Fig. \ref{fig:sum-all} compares the infections in Fig. \ref{fig:sum-no-0004}, Fig. \ref{fig:sum-hyst}, and Fig. \ref{fig:sum-rl}.


\section{Detailed Description} \label{sec:math}

We now proceed to give a detailed description of the mathematical model, for those interested in technical details. 

\subsection{Defining the network and population structure} \label{subsec:netdef}

The network data input needs to include, (i) the $(x,y)$ location of each node (in this case, district), (ii) the population of each node in the network, and (iii) optionally, the strength of connectivity between each pair of nodes in normal times. Note that the third input is optional because one can use gravitational models \cite{allamanis2012evolution} to approximate these connections. We define the population in node $n$ as $\rho_n$. We also generate the strength of connectivity $\kappa_{n1,n2}$ between each pair of nodes. The strength is defined to be directly proportional to the product of the populations, and inversely proportional to the square root of distances between the nodes. This last portion deviates from traditional gravitational models \cite{allamanis2012evolution}. However, the square root ensures that large districts remain strongly connected (for example, Mumbai and Delhi) even though their distances may be large.


We also need to model the circulation of people within nodes and to other nodes. The parameters of this process are listed in Table \ref{tab:population}. Instead of physically modelling the movement of individuals, we compute the number of external (other node) population exposed by,
\begin{equation}
POP_{ext,n1} = u_{n1,d}\sum_{n2\neq n1} \kappa_{n1,n2} \cdot u_{n2,d} \cdot \rho^*_{n2}.
\label{eq:popext}
\end{equation}
Note that $\rho^*_{n2}$ is the number of people in $n2$ who are \textit{not} showing symptoms (we assume that symptomatic people are not allowed to travel to other nodes), and $u_{n,d}$ is a binary indicator which is 1 if node $n$ is open, and 0 otherwise\footnote{This implies there is no external population exposure when $u_{n1,d}=0$.}. Similar to the above expression, we compute the number $\epsilon_{n2}$ of asymptomatic infected and exposed people from other nodes that will travel to $n1$,
\begin{equation}
INF_{ext,n1} = u_{n1,d}\sum_{n2\neq n1} \kappa_{n1,n2} \cdot u_{n2,d} \cdot \epsilon_{n2},
\label{eq:infext}
\end{equation}
The internal population exposure is given by,
\begin{align}
POP_{int,n1} = & u_{n1,d}\,(1-\psi_{lock})\,\left(\rho^*_{n1} + \psi_\mathrm{quar,leak}\,q_{n1}\right) \nonumber \\ & + \psi_\mathrm{lock}\left(\rho^*_{n1} + \psi_\mathrm{quar,leak}\,q_{n1}\right),
\label{eq:popint}
\end{align}
where $\psi_{lock}$ is the fraction of people that circulate within the node even when under lockdown. As before, $\rho_1^*$ is the number of people in $n1$ who are not showing symptoms, while $q_{n1}$ is the number of people who are showing symptoms (and should ideally be in quarantine). A similar computation gives us $INF_{int,n1}$, the number of infected people circulating within $n1$. On any given day, the probability of a susceptible to exposed state conversion for any person is given by,
\begin{equation*}
PROB_{exposure} = \frac{INF_{int,n1}+INF_{ext,n1}}{POP_{int,n1}+POP_{ext,n1}}.
\end{equation*}

\begin{table}
	\caption{Population circulation and contact parameters.}
	\label{tab:population}
	\begin{center}
		\begin{tabular}{|l|l|C{4.5cm}|}
			\hline
			Notation & Value & Explanation \\
			\hline
			- & 5000    & Number of people initially infected and randomly dispersed in network \\
		    \hline
$\lambda$ & $0\leq \lambda \leq 1$  & Fraction of population in a node, above which node must be locked \\
    \hline
$D_\mathrm{lock}$ & 7     & Number of days which for which a lockdown is compulsorily enforced \\
    \hline
$\psi_\mathrm{lock}$ & 0.01  & What fraction of the population comes into contact within the node on a given day, when node is locked down \\
 \hline
$\psi_\mathrm{quar,leak}$ & 0.1   & Fraction of symptomatic population that breaks quarantine and is at large \\
 \hline
$u_{n,d}$ & 0 or 1 & Whether a node $n$ is open$=1$ or locked$=0$ on day $d$ \\
 \hline
$\rho_n$ & $\rho \geq 0$ integer & Population of node $n$ (after subtracting any that are dead) \\
 \hline
$\kappa_{n1,n2}$ & $\kappa_{n1,n2} > 0$ & Connectivity strength between nodes $n1$ and $n2$ \\
			\hline
		\end{tabular}
	\end{center}
\end{table}

\subsection{Defining and modelling disease parameters}

\begin{figure}
    \centering
    \includegraphics[width=0.3\textwidth]{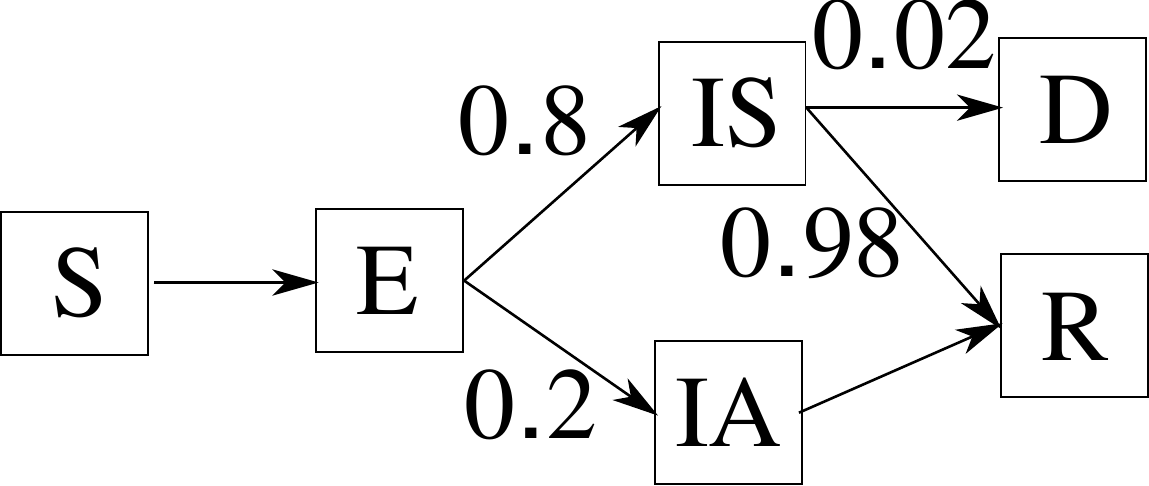}
    \caption{Change in state for each person. S is susceptible, E is exposed (virus in the body but not yet affecting the immune system), IS is infected (showing symptoms), IA is asymptomatic carrier, D is dead, and R is recovered. Note that numbers are capture probabilities and not rates of change.}
    \label{fig:states}
\end{figure}

The parameters of infectious diseases that we have captured in this model are listed in Table \ref{tab:disease}, though these can be changed based on demographics or new information about the disease. The stages of the disease that we consider are shown in Fig. \ref{fig:states}, based on prior literature \cite{perez2009agent} and known characteristics of Covid-19. We assume that people exposed to the disease will eventually show symptoms with a probability $P_\mathrm{E,IS} = 0.8$. The other 20\% of people will be asymptomatic carriers (state IA). All people in states E, IS, and IA are capable of transmission to others. People in the IS state die with a probability $P _\mathrm{IS,D} = 0.02$, while the rest go the Recovered state R. We assume that asymptomatic carriers do not die, but always go to R. The rate at which changes in state occur are inversely proportional to the stage durations as listed in Table \ref{tab:disease}. For example, the probability of an E state person moving to IS or AA is equal to $1/D_\mathrm{cub}$ on any given day. The number of such transitions on a day is produced by the requisite number of draws (equal to the population in a node in that state) from a multinomial distribution, with the relevant probabilities ($[1/D_\mathrm{cub},1-1/D_\mathrm{cub}]$ in the example given above). New exposures are produced by the circulation model as described previously. At this point (April 2020), we do not consider re-infections of recovered people, because it is unclear if and when this can happen. We can expand the model if required.

\begin{table}
	\caption{Properties of the disease.}
	\label{tab:disease}
	\begin{center}
		\begin{tabular}{|l|l|C{4.5cm}|}
			\hline
			Notation & Value & Explanation (days in average) \\
			\hline
			$D_\mathrm{cub}$ & 10    & Incubation days \\
			\hline
            $D_\mathrm{inf}$ & 14    & Days for which symptoms are shown \\
            \hline
            $D_\mathrm{trf}$ & 1     & Days after exposure when transmission can begin \\
            \hline
            $P_\mathrm{E,IS}$ & 0.8    & Probability of showing symptoms \\
            \hline
            $P_\mathrm{IS,D}$ & 0.02   & Prob. of symptomatic person dying \\
            \hline
            $P_\mathrm{trf}$ & 0.1     & Probability of infection of a susceptible person who comes into contact with an infected person \\
			\hline
		\end{tabular}
	\end{center}
\end{table}

\subsection{Reinforcement learning model}

\textbf{Deep RL architecture: }We use an RL algorithm known as Deep Q Network \cite{mnih2015human} to compute the optimal lockdown policies. At the beginning of each week, the model observes the relevant states for each node (described below) and produces a binary decision: $u_{n,d}=1$ for open and $u_{n,d}=0$ for lockdown. This decision is derived from a fully connected neural network.

We use a single network for computing the values of both actions, in each state. The input size is 8 (the number of features). There are 2 hidden layers with 12 and 8 neurons respectively, each with \textit{tanh} activation. The output layer has 2 \textit{linear} neurons, producing the long-term value estimates for either action. Once an action is computed for all nodes, the simulation rolls forward for 7 days (one day at a time) before calling the model for a new set of decisions.

\textbf{State space: }The RL algorithm uses the following input features to take the lockdown decision for each node once per week:
\begin{enumerate}
\item Population of own node,
\item Fraction infected (symptomatic) in own node,
\item Fraction infected (symptomatic) in overall population,
\item Fraction of population recovered in own node, 
\item Fraction dead in own node,
\item Potential infecting agents from rest of population,
\item Fraction increase in symptomatic population in own node in the last few days, and
\item Fraction increase in symptomatic population overall in the last few days.
\end{enumerate}

\textbf{Reward: }Each simulation lasts for 52 weeks (364 days). However, the terminal rewards are computed as soon as the number of active infections in the whole network goes to 0 (only S, R, D people: no further state changes are possible after this point), or when the 52-week window is completed. The terminal reward for node $n$ is given by,
\begin{equation*}
R_{n,\mathrm{term}} = a - \left( \frac{c_\mathrm{dead}\, x_{n,\mathrm{dead}}}{\rho_n} + \frac{c_\mathrm{inf}\,x_{n,\mathrm{inf}}}{\rho_n} + c_\mathrm{lock}\,d_{n,\mathrm{lock}} \right)
\end{equation*}
where $c_\mathrm{dead}=2.5\times 10^5$ is the penalty multiplier for number of deaths, $c_\mathrm{inf}=10^5$ is the penalty multiplier for number of infections, $c_\mathrm{lock}=\frac{1}{364}$ is the penalty multiplier for number of days of lockdown, $x_{n,\mathrm{dead}}$ is the final number of deaths in node $n$, $x_{n,\mathrm{inf}}$ is the final number of infections in node $n$, and $d_{n,\mathrm{lock}}$ is the number of days for which node $n$ was locked down. For an average node with a population of $2\times 10^6$, the cost equivalence is 1 week of lockdown for 0.4 infections or 0.15 deaths. The constant $a=2$ makes no difference to learning, and is only used for keeping the net reward close to 0 for illustration. Note that the terminal reward for each node is independent of the performance of other nodes. This value is backpropagated up to the start of the episode using a discount factor of $0.99$. We emphasize that all these cost parameters are user-modifiable, and will result in different learnt and optimized policies.

In addition to the terminal rewards, we also define step rewards as given by,
\begin{equation*}
R_{n,\mathrm{step}} = - \frac{c_\mathrm{dead}\Delta x_{n,\mathrm{dead}} + c_\mathrm{inf}\Delta x_{n,\mathrm{inf}} + c_\mathrm{lock}(1-u_{n,d})}{\rho_n},
\end{equation*}
where the $\Delta$ terms are changes in state over the past week, the last term penalises a lockdown, and the normalisation factor is the population of the node. 

\textbf{Training: }The target reward for each step is the average of the discounted terminal reward and the step reward, and is applied to the value output for the chosen action (0 or 1). The predicted reward for the other action is left untouched. The loss is computed using mean squared error (\textit{mse}). Mean-squared error is a standard loss function when trying to match the \textit{value} of an output. For a detailed discussion of loss functions, see \cite{Sutton:2012}. We roll out an entire episode before computing the rewards (the Monte-Carlo method \cite{mooney1997monte}), and thus we know the precise value of `correct' reward for each output. This is the value with respect to which we aim to minimise the prediction error (or loss). Training is carried out after each simulation using stochastic gradient descent optimiser in \textit{keras} with a learning rate of 0.001 and a momentum of 0.8, for 5 epochs. The collected memory is discarded after every simulation episode. During simulation, we choose a policy output (lock down or keep open) based on the predicted values for each outcome. If $v_0$ is the predicted value of action $0$ (lock down) in a given state, and $v_1$ is the predicted value of action $1$, then probability of choosing each action is,
\begin{align*}
    P(u_{n,d}=0) & = \frac{e^{\alpha v_0}}{e^{\alpha v_0}+e^{\alpha v_1}},\text{ and } \\
    P(u_{n,d}=1) & = \frac{e^{\alpha v_1}}{e^{\alpha v_0}+e^{\alpha v_1}}.
\end{align*}
The value of $\alpha$ results in a variety of standard methods in literature. Specifically, $\alpha=0$ implies that either action is taken with 0.5 probability (uniform random). A value of $\alpha=1$ results in the softmax distribution, while $\alpha=\infty$ results in purely greedy choice (always choose the action with higher value). In this study, we use $\alpha=5$, which is quite close to the greedy choice. However, both actions have a reasonable change of being picked if $v_0 \approx v_1$.

\begin{figure}
\centering
\includegraphics[width=0.45\textwidth]{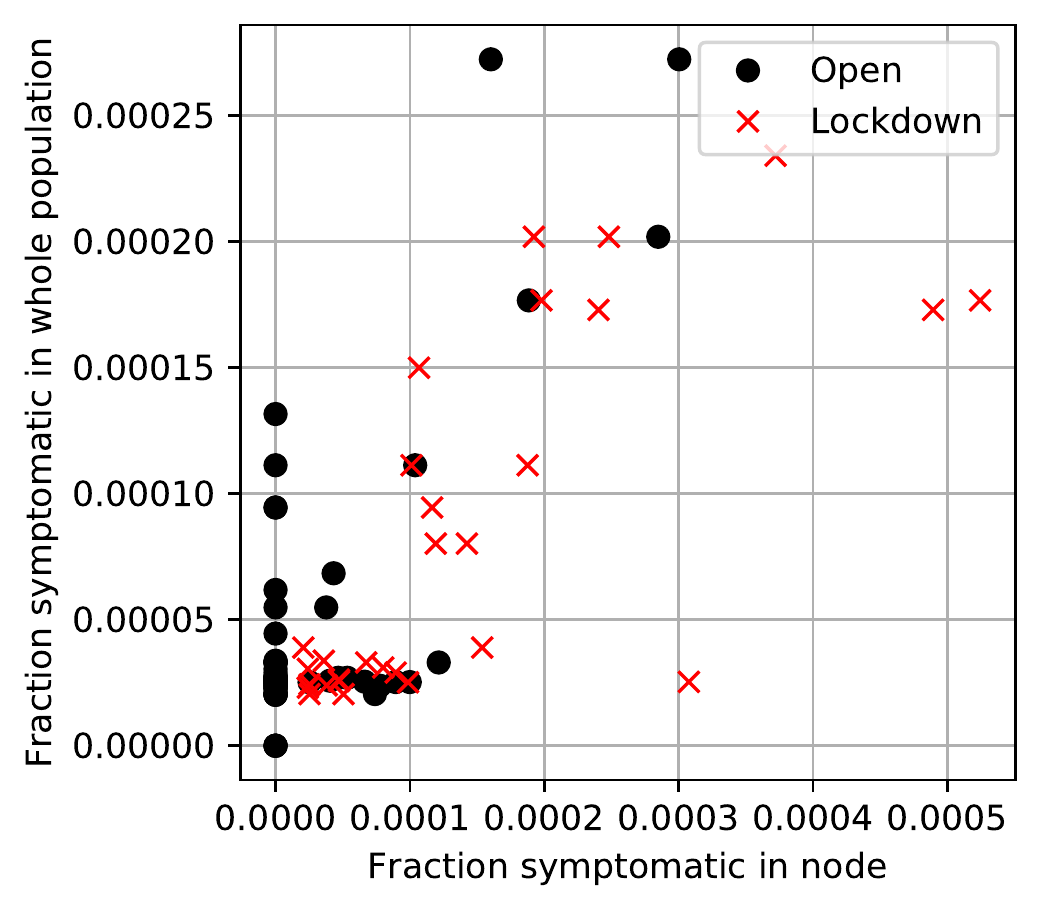}
\caption{Policy outcomes as a function of two features: Overall symptomatic population in the country, and the symptomatic population within a district.}
\label{fig:policy}
\end{figure}

\textbf{Analysis: }One of the claims that we made in the initial portion of the paper, was that our RL algorithm can be analysed, and is not a black box. One thread of analysis is to visualise the decisions taken by the algorithm as a function of the input features. Fig. \ref{fig:policy} shows such a plot with two features: the total symptomatic population in the country, and the symptomatic population within a district. We note that the decisions to keep the district open are concentrated towards the origin, where both features are close to 0. The policy is sensitive to increases along either dimension (as seen by the spread of lockdown decisions). However, it is more sensitive to the symptomatic population within the district (node). Note that the policy decides to keep the node open if its symptomatic population is 0, regardless of the level of infection in the rest of the country. It is clear that other features also affect the decisions (there is some mixing between the open and lock decisions). Further sophisticated visualisation of the policy is part of future work. 


\section{Further extensions} \label{sec:enhance}

We describe two types of extensions of this work. The first is an increase in the fineness of control, from a binary decision space to a more nuanced categorisation. The other is validation of this model against real-world data. While vital to the use of the proposed model as a decision support tool, the real-world data contains some unanswered questions. This is the reason we have not introduced validation as a central theme at this point in time (April 2020).

\subsection{Fine control of lockdown severity} \label{subsec:finecontrol}

\begin{figure}[t]
\centering
\includegraphics[width=0.48\textwidth]{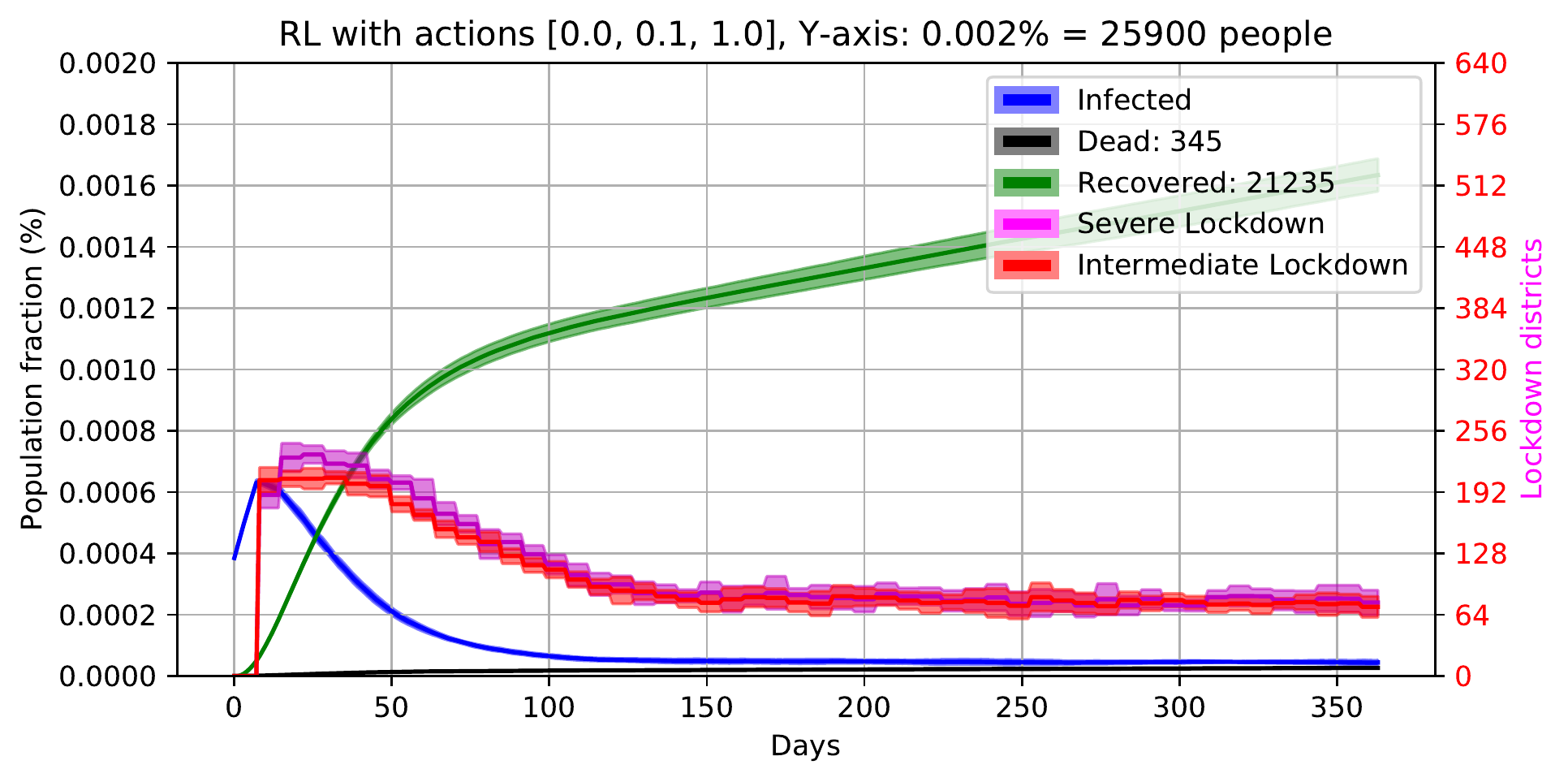}
\vskip-10pt
\caption{Simulated evolution of epidemic in India under a reinforcement learning policy with three possible actions.}
\label{fig:sum-multiact}
\vskip-10pt
\end{figure}

One easily identifiable drawback in the proposed RL model, is the binary nature of lockdown decisions. Observation of policies around the world show a spectrum of decisions, from social distancing, school closures, industry closures, to full lockdowns. In order to address this aspect, we can allow the decisions $u_{n,d}$ to take one a discrete set of values. Note that this allows us to effectively control the number of people circulating across nodes as well as within nodes, by modifying $u_{n1,d}$ and $u_{n2,d}$ in equations (\ref{eq:popext}), (\ref{eq:infext}) and (\ref{eq:popint}). In Fig. \ref{fig:sum-multiact}, we show the evolution of the epidemic under a 3-action regime: (i) no lockdown, (ii) 90\% lockdown, and (iii) 100\% lockdown. While the RL algorithm takes more episodes to optimise the policy with this higher level of freedom, we note that it is able to achieve similar results with less severe lockdowns (while 100\% lockdowns were introduced in 380 districts in Fig. \ref{fig:sum-rl}, this is only necessary in about 250 districts in Fig. \ref{fig:sum-multiact}).

\begin{figure}
\centering
\vskip-20pt
\includegraphics[width=0.48\textwidth]{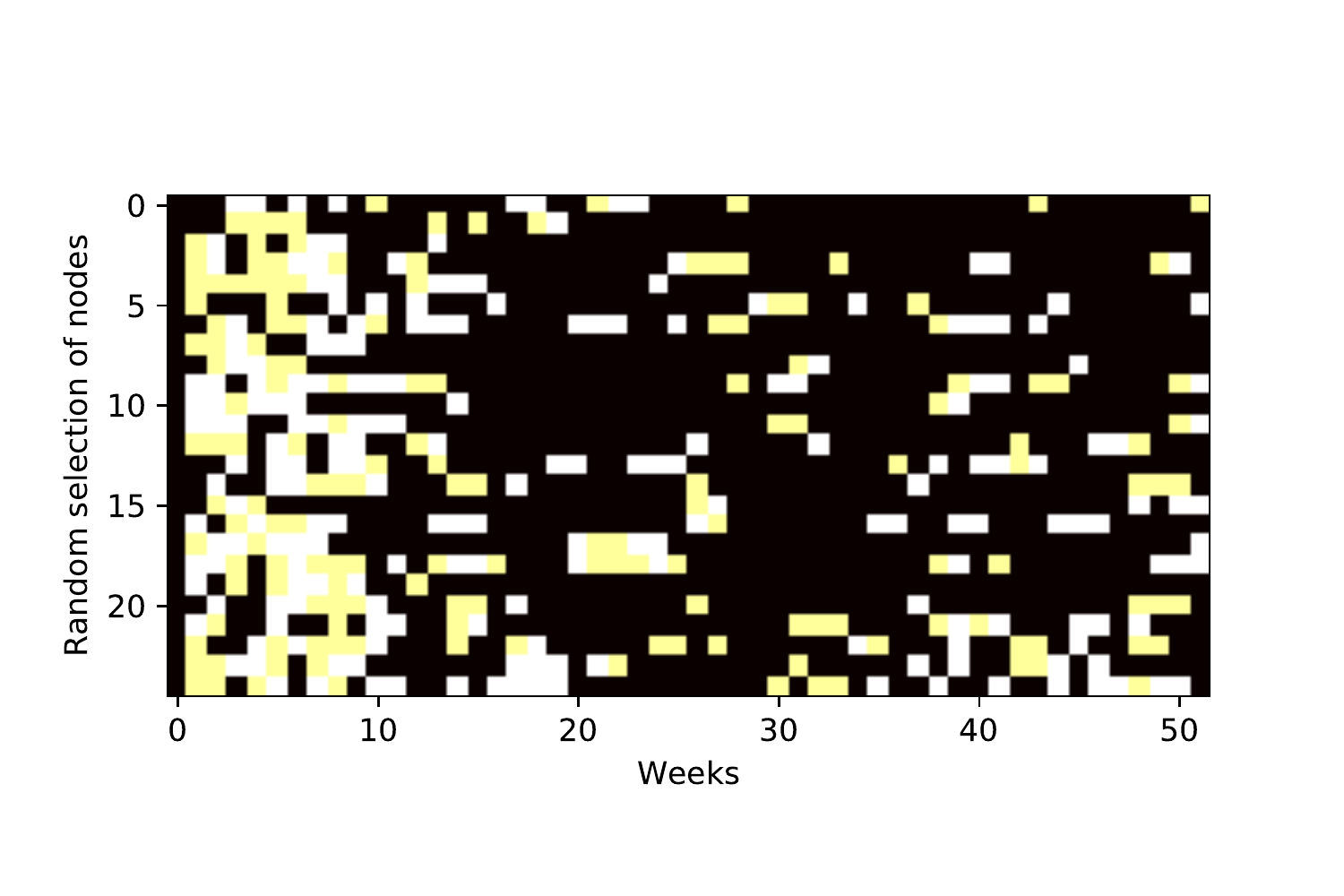}
\vskip-20pt
\caption{Simulated lockdown histories for a subset of 25 districts, over 52 weeks. Yellow pixels indicate 90\% lockdown in that district and week, white pixels indicate 100\% lockdown, and black pixels indicate no lockdown.}
\label{fig:lock-multiact}
\vskip-20pt
\end{figure}

Fig. \ref{fig:lock-multiact} illustrates the policy effect in a subset of 25 districts, over the course of 52 weeks. Yellow pixels indicate 90\% lockdown in that district and week, white pixels indicate 100\% lockdown, and black pixels indicate no lockdown. We note that most lockdowns are concentrated at the beginning, towards the left of the figure. Later in the year, lockdowns are intermittently applied as needed. Furthermore, the switches are mostly smooth from open to 90\% lockdown to 100\% lockdown, and vice versa.

The algorithm is capable of working with an arbitrary number of such categorical actions, as long as the underlying network model is capable of interpreting and applying the choices effectively. Furthermore, it is not necessary for each action to mean a specific numerical value for $u_{n,d}$. Each enumerated action could imply a specific type of policy, specified by the opening or closing of various activities. For example, one output could mean `keep industries open but close schools and malls', while another could implement only mild social distancing. The DQN approach as described earlier, is natively capable of handling categorical outputs.

\subsection{Fitting the model to real-world data} \label{subsec:validation}

\begin{figure*}
\includegraphics[width=0.99\textwidth]{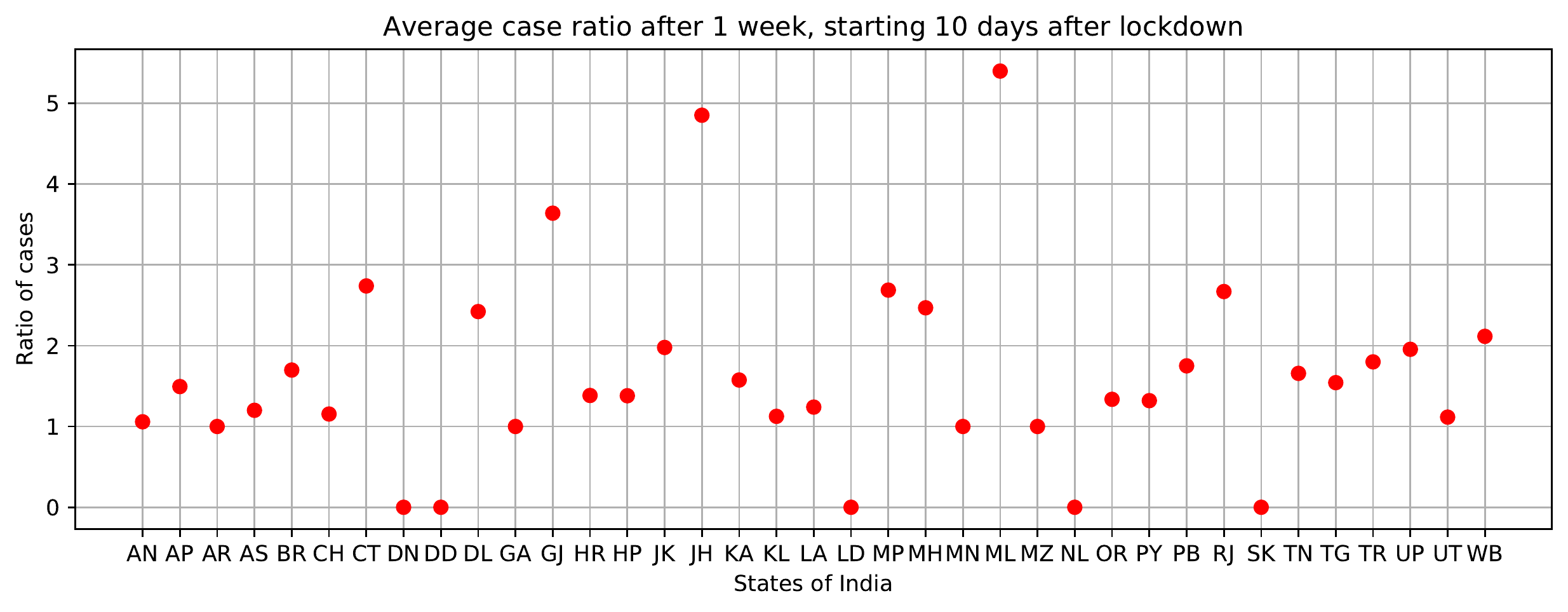}
\caption{Ratio of active cases in each state in India with a gap of 1 week, starting 10 days after the lockdown was implemented on March 25, 2020.}
\label{fig:ratios}
\end{figure*}
\begin{figure*}
\centering
\includegraphics[width=0.7\textwidth]{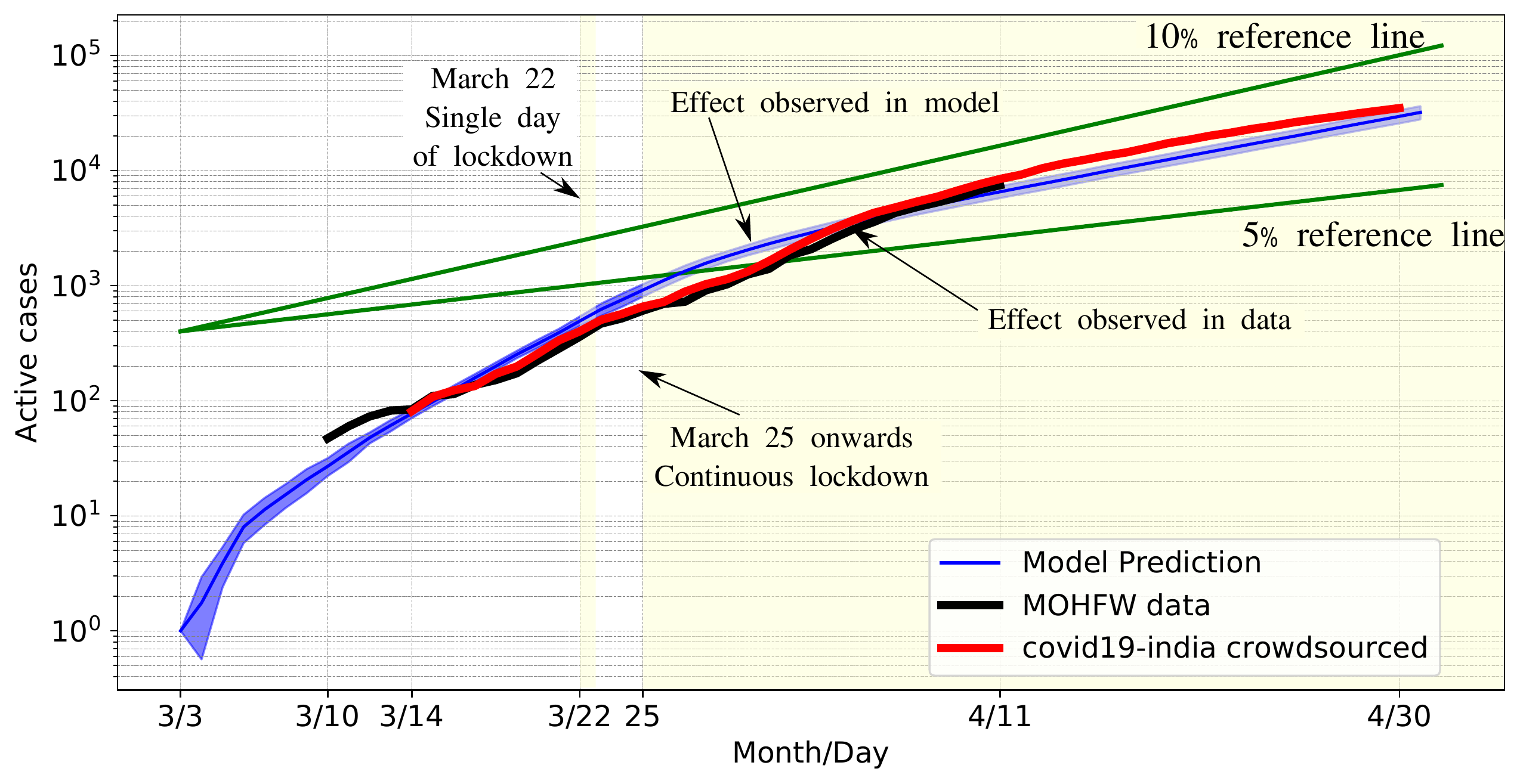}
\caption{Comparison of model output with public data.}
\label{fig:val-49}
\end{figure*}
\begin{figure*}
\centering
\includegraphics[width=0.7\textwidth]{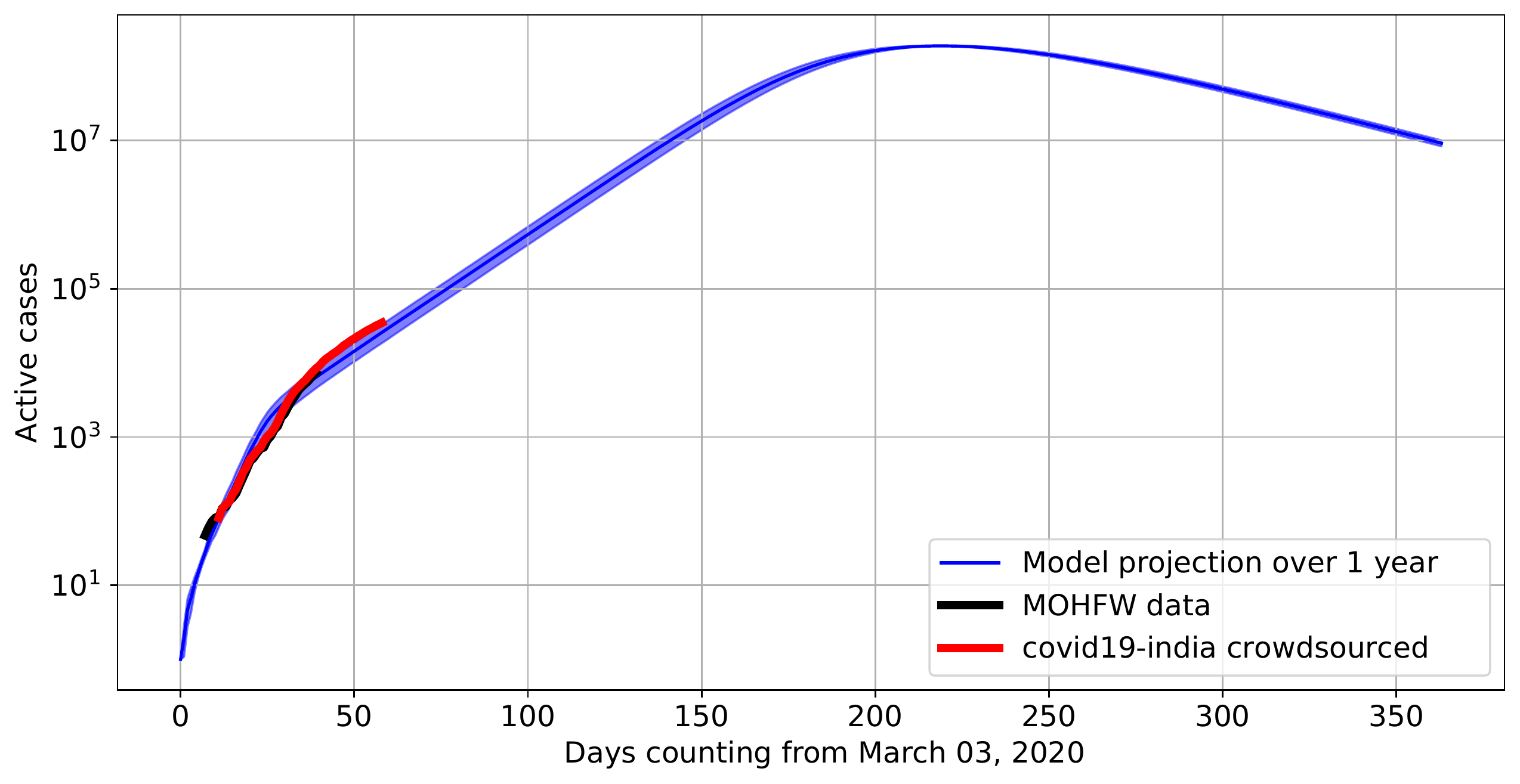}
\caption{Projected model output for 1 year starting March 03, 2020, after fitting to available public data. \textbf{This model contains gross approximations, and should not be construed as an accurate prediction of evolution of the disease.}}
\label{fig:val-364}
\end{figure*}

One of the critical measures of the usefulness of a model is how well it explains real-world data. In this section, we use data from two public sources\footnote{https://www.mohfw.gov.in/ retrieved April 12, 2020}$^,$\footnote{http://api.covid19india.org/states\_daily\_csv/confirmed.csv retrieved May 01, 2020} which are in reasonably good agreement with each other. Both data sets are available at the level of `states' in India, each of which contains several districts. The first data set is from the Ministry of Health, with our version containing counts from March 10 to April 11. The second is a crowd-sourced data set from March 14 to April 30 (latest day as of completing this manuscript). Since these data sets contain only known positive cases, we make a few modelling assumptions:
\begin{itemize}
\item We assume that on any given day, only 75\% of symptomatic cases and 25\% of asymptomatic cases are captured in the data.
\item We assume that the first known cases on March 10 started as `Exposed' cases on March 03 (7 days in advance, the average incubation period of Covid-19). 
\item The initial cases as available per state, are assigned to the capital districts of that state.
\end{itemize}
We then proceed to simulate the model with the actual interventions imposed in India. These include a single day of lockdown on March 22, followed by a full lockdown from March 25 onwards. The efficiency of lockdown for each state is derived from the same data sets. We note from Fig. \ref{fig:ratios} that different states show different rates of growth, even when the effects of lockdown have stabilised (more than 10 days after the imposition on March 25). The efficiency of lockdown (value of $u_{n,d}$) for all districts in each state is assumed to be inversely proportional to this growth ratio. A simulation of our network model with these efficiency numbers, starting from the assumed initial state on March 03, is shown in Fig. \ref{fig:val-49}. We note that the fit is reasonably good. 

However, a continued simulation of the same lockdown for a period of one year (shown in Fig. \ref{fig:val-364}) shows the cases increasing to over 1 million before starting to fall. This is an artefact of the relatively mild reduction in slope after the lockdown was imposed (approximately 5\% per day as of May 01, 2020). Most models with moderate levels of lockdown efficiency show a sharper decrease in the number of cases after imposition, as compared to the observed data. We speculate that the observations may be skewed by the following factors:
\begin{itemize}
\item While the actual case growth may have slowed after lockdown, the testing rates are still catching up.
\item Our model does not explicitly account for individual events of large groups of people breaking quarantine, though we do incorporate an individual's probability of breaking quarantine. These violations may cause spikes in case counts even during the lockdown.
\item The number of false positives in the current tests is unknown, as is the type of test administered.
\item Apart from tests carried out per capita, the fraction of positive tests also varies significantly for different states. These differences could be caused by the type of test, number of false negatives or false positives, and the efficiency of contact tracing.
\end{itemize}
Furthermore, the long-term projections contain gross approximations because of (i) simplified network modelling, (ii) simplified epidemiological modelling, (iii) lack of consideration for external factors such as weather effects and lockdown relaxations. Therefore, we make no claims of accuracy regarding Fig. \ref{fig:val-364}, and instead present this as a challenge for accurate modelling.

\textbf{Future work:}
We plan to continue on improvements to the model, including parameter tuning using new data, and the incorporation of healthcare capacity in the framework. The latter can significantly affect mortality rates and hence economic outcomes. Furthermore, the optimal policy can be tuned specifically for each node, depending on the predominant type of economic activity and population characteristics.




\bibliographystyle{IEEEtran}
\bibliography{refs}

\end{document}